\documentstyle[aps,prl,multicol,epsfig]{revtex}
\begin{document}

\title{Superconducting gap and pair breaking in CeRu$_{2}$
studied by point contacts}

\author{A. V. Moskalenko, Yu. G. Naidyuk, and I. K. Yanson}

\address{$^1$B.Verkin Institute for Low Temperature Physics and Engineering,
National Academy  of Sciences of Ukraine, \\ 47 Lenin Ave., 61103,
 Kharkiv, Ukraine}

\author{M. Hedo, Y. Inada, and Y.$\bar{O}$nuki}

\address{$^2$Graduate School of Science, Osaka University, Toyonaka
560-0043, Japan}

\author{Y. Haga, and E. Yamamoto}

\address{$^3$Advanced Science Research Center, Atomic Energy Research
Institute of Japan, Tokai, Ibaraki 319-1195, Japan}

\date{\today}
\maketitle

\begin{abstract}
The superconducting gap in a CeRu$_{2}$ single crystal is
investigated by point contacts. BCS-like behavior of the gap
$\Delta $ in the temperature range below T$_{c}^{*}<$T$_{c}$,
where T$_{c}$ is the critical temperature, is established,
indicating the presence of a gapless superconductivity region
(between T$_{c}^{*}$ and T$_{c}$). The pair-breaking effect of
paramagnetic impurities, supposedly Ce ions, is taken into
consideration using the Scalski-Betbeder-Matibet-Weiss approach
based on Abrikosov-Gorkov theory. It allows us to recalculate the
superconducting order parameter $\Delta ^{\alpha} $ (in the
presence of paramagnetic impurities) and the gap $\Delta ^{P}$ (in
the pure case) for the single crystal and for the previously
studied polycrystalline CeRu$_{2}$. The value 2$\Delta^{P}$(0)$
\approx $2 meV, with 2$\Delta ^{P}$(0)$/$k$_{B}$T$_{c} \approx
$3.75, is found in both cases, indicating that CeRu$_{2}$ is a
``moderate'' strong-coupling superconductor.
\end{abstract}



\begin{multicols}{2}
\section*{Introduction}

The superconducting gap $\Delta$ of CeRu$_{2}$ has been evaluated
by point-contact Shottky tunneling (PCT) \cite{1}, break-junction
tunneling (BJT) \cite{2}, point contact spectroscopy (PCS)
\cite{3}, and STM experiments \cite{4}. The necessity of new PCS
experiments is due to the discrepancy between the results obtained
by different experimental methods. From the PCT measurements
2$\Delta $(0)/k$_B$T$_{c}$ is estimated as 6.6$\pm $0.6, while the
BJT experiments yielded 2$ \Delta $(0)/k$_B$T$_{c}$=4.4. These
values are remarkably larger than our previous PCS result 3.1$ \pm
$0.1 \cite{3}, which is more consistent with the recent tunneling
data 2$\Delta $(0)/k$_B$T$_{c}$=3.3 \cite{4}. The measurements of
the superconducting gap were performed by different methods and on
samples of different quality. In this paper we present a
comparison of the superconducting gap behavior of samples with
different quality studied by one method. We also propose a
procedure for $\Delta $ correction based on taking pair-breaking
effect into account, which results in almost equal gap values for
both samples.\\

\section*{Experiment and results}

We have studied the superconducting gap in single-crystal
CeRu$_{2}$ samples by measuring \textit{dV/dI} for S-c-N (here S -
is a superconductor, c - is a constriction and N - is a normal
metal) point-contacts. The single crystal was grown by the
Czochralski pulling method in a tetra-arc furnace. Its residual
resistivity ratio (RRR) is 120, residual resistivity
$\rho_{0}=1\mu \Omega $cm and T$_{c}$=6.3\,K. The polycrystalline
CeRu$_{2}$  studied in \cite{3} had RRR=14,
$\rho_{0}=31.5\,\mu\Omega\,$cm and T$_{c} $=6.2 K, that is, it had
much lower quality. The point-contact characteristics presented
were obtained on the cleaved surface of CeRu$_{2}$ for both
samples. The sample size was about 1x1x5 mm$^{3}$. The sample were
cleaved in air at room temperature. The PCs were prepared by
touching this surface with the edge of an Ag or Cu
counterelectrode, which were cleaned by chemical polishing. The
experimental cell with the sample and counterelectrode was
immersed directly in liquid $^{4}$He to ensure good thermal
coupling. The measurements were carried out in the temperature
range 1.7-6.7 K. The differential resistance \textit{dV/dI} of the
PCs was recorded versus the bias voltage using a standard lock-in
amplifier technique, modulating the direct current \textit{I} with
a small 480 Hz ac component.

The Blonder, Tinkham and Klapwijk (BTK) theory \cite{5} is
commonly used to describe the behavior of the current-voltage
characteristics of clean S-c-N microconstrictions. As in our
previous publication [3], here we have used this model, which
takes into account the Andreev reflection on the S-N interface
\cite{5}, to fit the measured \textit{dV/dI(V)} curves of PCs.
According to the theory \cite{5} a maximum at zero-bias voltage
and a double-minimum structure around \textit{V}$\sim \pm\Delta
/$e on the \textit{dV/dI} curves manifests the Andreev reflection
process with a finite barrier strength parameter \textit{Z}. It
follows from the equations for the current-voltage characteristics
\begin{eqnarray}\label{BTK}
I(V) &\sim &\int_{-\infty }^{\infty}
T(\epsilon)\left(f(\epsilon-{\rm e}V)-f(\epsilon)\right) {\rm
d}\epsilon,  \\ T(\epsilon) &=&\frac{2\Delta
^2}{\epsilon^2+(\Delta
^2-\epsilon^2)(2Z^2+1)^2},~~~|\epsilon|<\Delta \nonumber \\
T(\epsilon)
&=&\frac{2|\epsilon|}{|\epsilon|+\sqrt{\epsilon^2-\Delta
^2}(2Z^2+1)},~~~ |\epsilon|>\Delta ~, \nonumber
\end{eqnarray}
where f($\epsilon$) is the Fermi distribution function. The
broadening of the quasiparticle density of states N($\epsilon,
\Gamma $) in the superconductor was taken into account according
to Dynes et al. \cite{6}:
\begin{equation}
N(\epsilon,\Gamma )={\rm Re}\left\{ \frac{\epsilon-i\Gamma }
{\sqrt{(\epsilon-i\Gamma)^2-\Delta ^2}}\right\}, \label{Dynes}
\end{equation}
where $\Gamma $ is the broadening parameter.

\begin{figure}[t]
\begin{center}
\includegraphics[width=8.5cm]{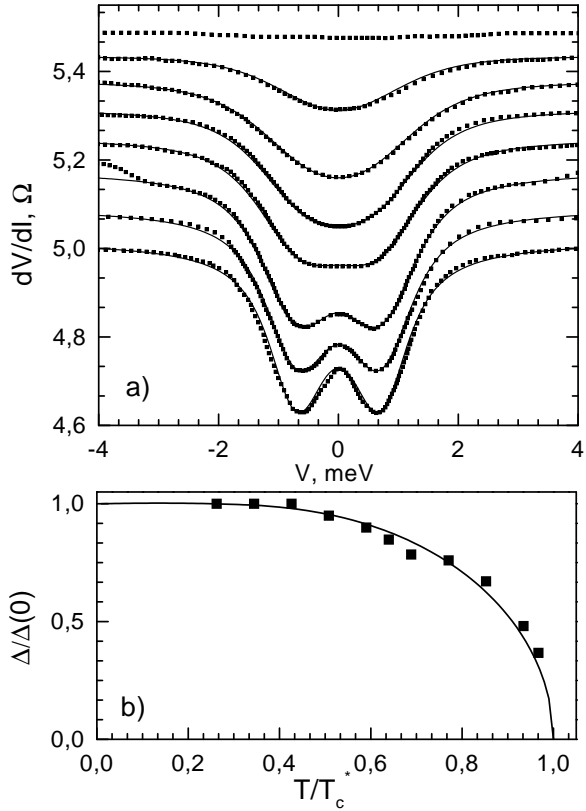}
\end{center}
\caption[]{a) Temperature dependence of the experimental
d$V$/d$I(V)$ curves (squares) for a CeRu$_{2}$-Ag point contact
with R$_{N}$=5$\Omega $ along with the fit using Eqs.(1),(2) with
$\Gamma \cong $0.13 meV and Z=0.43 (solid lines). The curves are
shifted vertically for clarity. The temperature from the bottom
curve to the top curve is: 1.6, 2.1, 2.6, 3.6, 4.2, 5.2, 5.9, 6.3
K. b). Temperature dependence of the superconducting gap $\Delta $
extracted from the fit in Fig.1,a. $\Delta $(0)=0.79 meV and
T$_{c}^{*}$=6.1K, with 2$\Delta (0)/$k$_{B}$T$_{c}^{*}$=3.05. The
solid line is the BCS curve.}
\end{figure}
In Fig.1,a a series of experimental \textit{dV/dI(V)} curves of
PCs based on the CeRu$_{2}$ single crystal are presented along
with the fitted ones for different temperatures. The  good
agreement between experimental and theoretical curves allowed us
precisely to determine $\Delta $ along with its temperature
dependences from calculations according to (1), (2). The average
value of the gap $\Delta$ of the CeRu$_{2}$ single crystal
extracted from the fit for 5 PCs is (0.83$ \pm $0.07) meV, with
2$\Delta (0)/$k$_{B}$T$_{c}^{*}$=3.23$ \pm $0.23 and T$_{c}^{*}$=
(5.9$ \pm  $0.2) K. The maximum $\Delta $(0) was 0.95 meV and
T$_{c}^{*}$=6.1 K, where T$_{c}^{*}$ is the extrapolated
temperature at which the gap drops to zero (see Fig.1,b). The
temperature dependence of the superconducting gap extracted from
the curves on Fig.1,a is presented on Fig.1,b and has a BCS-like
behavior as in the polycrystalline sample \cite{3}. The
extrapolated critical temperature T$_{c}^{*}$ for single crystal
is higher and the gapless region is smaller than in the
polycrystalline sample \cite{3}. The gap value (averaged for 5 PCs
as well) grew from (0.51$ \pm $0.07) meV for the polycrystal to
the value indicated above for the single crystal of CeRu$_{2}$.
This has a natural explanation considering the difference in the
quality of the samples. The contacts made on the more perfect
single-crystal CeRu$_{2}$ exhibited better superconducting
properties than those with the polycrystal.

\begin{figure}[t]
\begin{center}
\includegraphics[width=9cm]{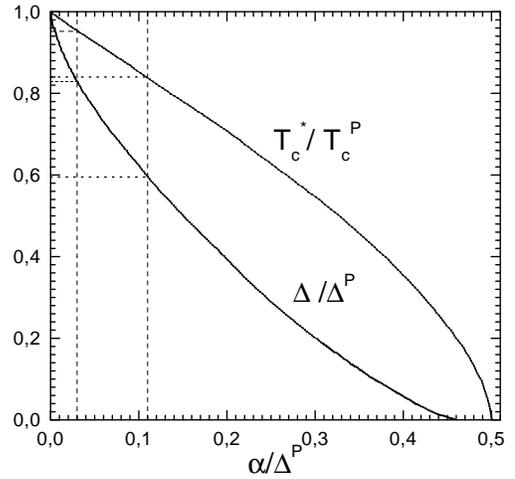}
\end{center}
\caption[]{ The T$_{c}^{*}$/T$_{c}^{P}$ and $\Delta (0)/\Delta
^{P}$(0) vs. $\alpha /\Delta ^{P}$(0) curves from Ref.\,8. The
dashed vertical lines indicate $\alpha /\Delta
^{P}$(0)=$\beta=$0.03 for the single crystal from Fig.1 and
$\beta=$0.11 for the polycrystal from Fig.1 in Ref.3 as determined
using the experimental values of T$_{c}^{*}$/T$_{c}^{P} $(dotted
horizontal lines). Dashed horizontal lines show the
$\Delta(0)/\Delta^{P}$(0)=$\gamma$ values determined.}
\end{figure}

In our previous paper \cite{3} the presence of a region of gapless
superconductivity in CeRu$_{2}$ between T$_{c}^{*}$ and T$_{c}$
was proposed to explain why T$_{c}^{*} \ne $ T$_{c}$. The gap was
assumed to be suppressed by the local magnetic moments, presumably
Ce, distributed randomly in the contact region. That is, because
of the lower purity (quality) of the polycrystal some of the
Ce-ions could be impurities.

The well known Abrikosov-Gorkov (AG) theory of a superconductor
containing paramagnetic (PM) impurities \cite{7} was considered
for explaining a gapless state in CeRu$_{2}$. The theory describes
a situation when in the presence of PM impurities the gap $\Delta
$ in the excitation energy spectrum drops to zero at a transition
temperature T$_{c}^{*}$, although the material is still a
superconductor in the sense of having pair correlations. The
transition temperature T$_{c}^{*}$ is lower than the critical
temperature T$_{c}$, and a range of temperatures between
T$_{c}^{*}$ and T$_{c}$ where $\Delta  $ is zero for any value of
the impurity concentration exists. The
Scalski-Betbeder-Matibet-Weiss (SBMW) - approach \cite{8} based on
AG-theory allows us to take into account a pair breaking caused by
spin-exchange scattering. As a measure of this effect produced by
PM impurities the inverse collision time for exchange scattering
$\alpha =\hbar/\tau_{ex}$ was used. The advantage of the SBMW
approach is the natural way in which the distinction between the
energy gap $\Delta $ and the order parameter $\Delta^{\alpha}$
arises when the effect of PM impurities on the density of states
is taken into account. The SBMW theory allows us to calculate the
order parameter $\Delta^{\alpha}$ of a superconductor with PM
impurities by transformation of the original expression (4.8) from
\cite{8}: $$\Delta\left( {T,\alpha} \right) = \Delta ^{\alpha}
\left( {T,\alpha} \right)\left[ {1 - \left( {\frac{{\alpha}
}{{\Delta ^{\alpha} \left( {T,\alpha} \right)}}} \right)^{2/3}}
\right]^{3/2}$$ into the following form:
\begin{equation}
 \Delta ^{\alpha} \left( {T,\alpha}  \right) = \Delta \left( {T,\alpha}
\right)\left[ {1 + \left( {\frac{{\alpha} }{{\Delta \left(
{T,\alpha} \right)}}} \right)^{2/3}} \right]^{3/2}.
\end{equation}

In (3) the pair-breaking parameter $\alpha $ is unknown. It was
determined from the T$_{c}^{*}$/T$_{c}^{P}$ and $\Delta (0)/\Delta
^{P}$(0) versus $\alpha /\Delta ^{P}$(0) curves (the superscript P
indicates a pure superconductor; we also suppose that T$_{c}^{P
}\equiv $T$_{c}$) shown in Fig.2. The value of T$_{c}^{*}$ was
taken from the experimental temperature dependence of $\Delta $ as
the value extrapolated  according to the BCS theoretical curve
(see Fig. 1b). Then $\alpha /\Delta ^{P}$(0)=$\beta $
corresponding to T$_{c}^{*}$/T$_{c}^{P}$ was determined for the
particular point contact, and $\Delta (0)/\Delta ^{P}$(0)=$\gamma
$ at the value of $\alpha /\Delta^{P}$(0)=$\beta $ determined was
specified. The $\Delta (0)$ value was taken from a fit of the
experimental curve. Thus we obtained $\Delta^{P}$(0)=$\Delta
(0)/\gamma $ and, hence, $\alpha $=$\Delta ^{P}$(0)$\beta
$=$\Delta (0)\beta /\gamma $. The order parameter $\Delta
^{\alpha} (0) $was found from (3) to be (0.99$ \pm $0.05) meV for
the single crystal and (0.87$ \pm  $0.1) meV for the polycrystal.
Fig.3 shows the results of the calculations of $\Delta ^{\alpha}
({\rm T})$ from (3). The temperature dependences of the parameter
$\Delta^{\alpha }$ both for the purer sample and for the less
perfect one have behavior close to the BCS curve.

The order parameter $\Delta ^{P}$(0) of the pure superconductor
can be determined from Fig.2 using the value of $\gamma$ or
calculated from the expression (3.5) of Ref.\,8:
 $$\ln\left( {\frac{{\Delta
^{\alpha} \left( {0,\alpha}  \right)}}{{\Delta ^{P}\left( {0}
\right)}}} \right) = - \frac{{\pi} }{{4}}\frac{{\alpha} }{{\Delta
^{\alpha} \left( {0,\alpha} \right)},}$$
 which transforms into
\[
\Delta ^{P}\left( {0} \right) = \Delta ^{\alpha} \left( {0,\alpha}
\right)\exp \left( \frac{\pi}{4}\,\frac{\alpha}{\Delta ^{\alpha}
\left({0,\alpha}\right)}\right).
\]

The value of $\Delta^{P}$(0) has less scatter in comparison with
$\Delta^{\alpha}(0)$ and is about (1.02$\pm$0.05) meV with
2$\Delta^{P}(0)/$k$_{B}$T$_{c}$=3.8$\pm$0.2 for the single crystal
and $\Delta^{P}$(0)= (0.99$\pm$0.13)meV with 2$\Delta
^{P}(0)/$k$_{B}$T$_{c}$=3.7$\pm$0.5 for the polycrystal.

\begin{figure}[t]
\begin{center}
\includegraphics[width=9cm]{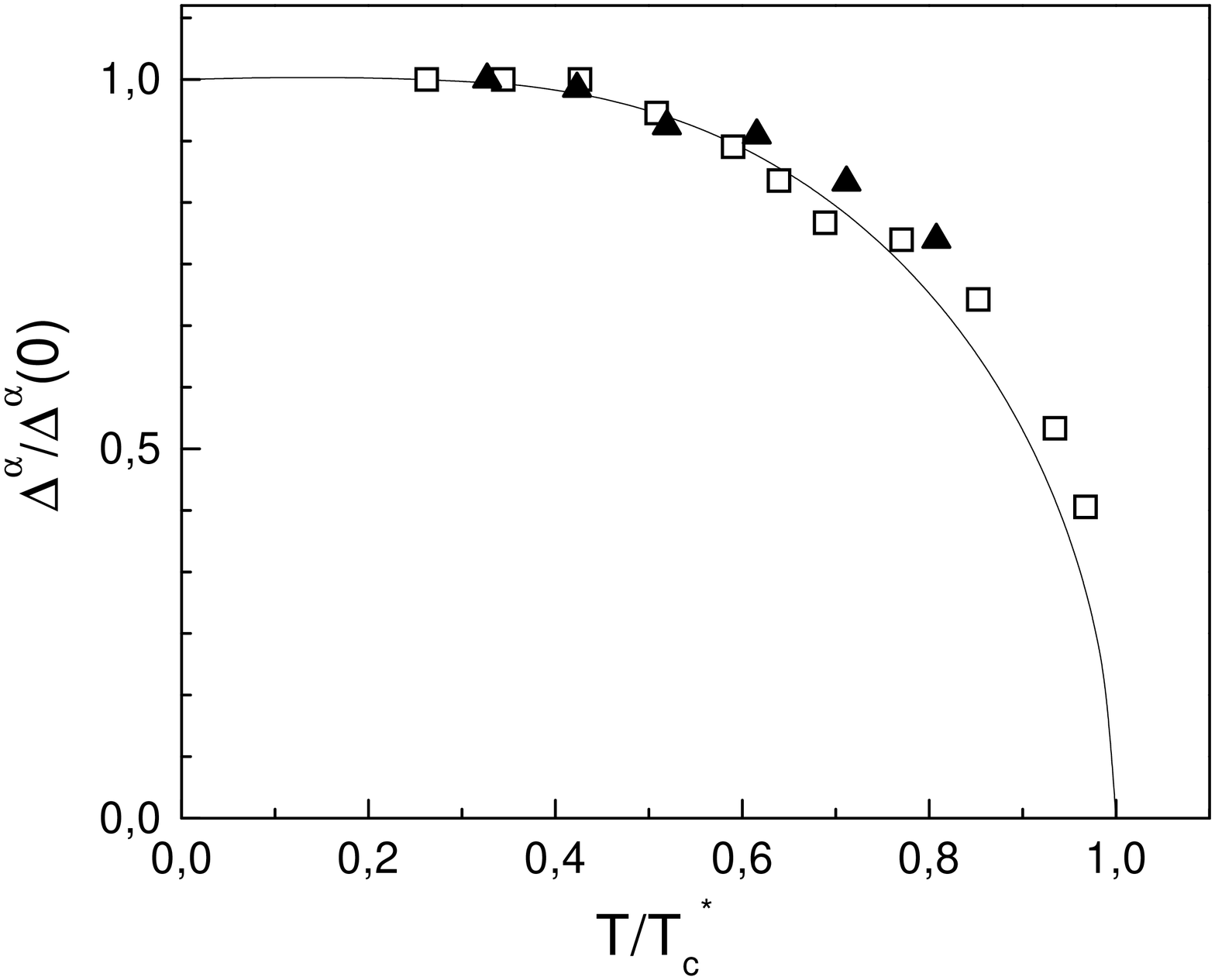}
\end{center}
\caption[]{Temperature dependences of the order parameter $\Delta
^{\alpha }$(T) for the single-crystal sample from Fig.1 (open
squares) and for the polycrystalline sample from Fig.1. in Ref.\,3
(triangles).}
\end{figure}

\section*{Discussion and conclusions}

As was shown earlier \cite{3,4} and in this paper the temperature
dependence of $\Delta$ in CeRu$_{2}$ has a BCS-like behavior, but,
with a lower critical temperature T$_{c}^{*}$. Because of the
difference in T$_{c}^{*}$ for the samples of different quality we
can conclude that in the cleaner one the influence of impurities
on the superconductivity is also weaker. Pair-breaking effects in
the contact area can be caused by the randomly distributed local
magnetic moments. It was noted by Joseph et al. \cite{9} that a
Ce-rich solid solution is present as a second phase in CeRu$_{2}$
in an amount up to 10\% in the samples of low quality. This means
that pair-breaking effects and gapless superconductivity in the
compound more probably are connected with the influence of
Ce-impurities. Calculations based on the SBMW approach gave very
close values of $\Delta^{P}$(0) and 2$\Delta
^{P}(0)/$k$_{B}$T$_{c}$ for poly- and single crystals. This
supports our assumptions about the influence of paramagnetic
impurities on superconductivity in CeRu$_{2}$ and gives a method
of $\Delta$ correction. This method of recovering the
superconducting parameters from point-contact \textit{dV/dI(V)}
characteristics can theoretically be improved by including in the
BTK fit a density of states modified by the pair-breaking effect.

\section*{Acknowledgement}
The investigations were carried out in part with the help of
donated by Alexander von Humbold Stiftung (Germany) equipment.

\end{multicols}
\end{document}